\documentclass[preprint,showpacs,preprintnumbers,amsmath,amssymb]{revtex4}

\newcommand\rf[1]{(\ref{eq:#1})}
\newcommand\lab[1]{\label{eq:#1}}
\newcommand\nonu{\nonumber}
\newcommand\br{\begin{eqnarray}}
\newcommand\er{\end{eqnarray}}
\newcommand\be{\begin{equation}}
\newcommand\ee{\end{equation}}

\newcommand\lb{\lbrack}
\newcommand\rb{\rbrack}

\newcommand\llb{\left\lbrack}
\newcommand\rrb{\right\rbrack}

\renewcommand\({\left(}
\renewcommand\){\right)}
\renewcommand\v{\vert}                     
\newcommand\bgv{\bigg\vert}              

\newcommand\bc{\begin{center}}
\newcommand\ec{\end{center}}

\newcommand\Tr{\mathop{\mathrm Tr}}                  
\newcommand\partder[2]{\frac{{\partial {#1}}}{{\partial {#2}}}}

\renewcommand\a{\alpha}

\renewcommand\d{\delta}

\newcommand\vareps{\varepsilon}
\newcommand\g{\gamma}
\newcommand\G{\Gamma}

\newcommand\h{\frac{1}{2}}
\renewcommand\k{\kappa}
\renewcommand\l{\lambda}
\renewcommand\L{\Lambda}
\newcommand\m{\mu}
\newcommand\n{\nu}
\renewcommand\o{\over}
\newcommand\om{\omega}

\newcommand\p{\phi}
\newcommand\vp{\varphi}
\renewcommand\P{\Phi}
\newcommand\pa{\partial}

\newcommand\pr{\prime}

\renewcommand\r{\rho}
\newcommand\s{\sigma}
\renewcommand\S{\Sigma}
\renewcommand\t{\tau}
\renewcommand\th{\theta}

\newcommand\wti{\widetilde}
\newcommand\twomat[4]{\left(\begin{array}{cc}  
{#1} & {#2} \\ {#3} & {#4} \end{array} \right)}

\newcommand\cA{{\mathcal A}}

\newcommand\cF{{\mathcal F}}

\newcommand\cL{{\mathcal L}}
\newcommand\cM{{\mathcal M}}

\newcommand{\ct}[1]{\cite{#1}}
\newcommand{\bib}[1]{\bibitem{#1}}
\newcommand\NPB[3]{\textsl{Nucl. Phys.} \textbf{B#1}, #3 (#2)}

\newcommand\PRD[3]{\textsl{Phys. Rev.} \textbf{D#1}, #3 (#2)}

\newcommand\PLB[3]{\textsl{Phys. Lett.} \textbf{#1B}, #3 (#2)}
\newcommand\CQG[3]{\textsl{Class. Quantum Grav.} \textbf{#1}, #3 (#2)}

\newcommand\PR[3]{\textsl{Phys. Reports} \textbf{#1}, #3 (#2)}

\newcommand\IJMPA[3]{\textsl{Int. J. Mod. Phys.} \textbf{A#1}, #3 (#2)}

\newcommand\MPLA[3]{\textsl{Mod. Phys. Lett.} \textbf{A#1}, #3 (#2)}

\begin{document}

\preprint{hep-th/0507193}

\title{Weyl-Conformally-Invariant Lightlike $p$-Brane Theories:\\
New Aspects in Black Hole Physics and Kaluza-Klein Dynamics}

\author{E.I. Guendelman and A. Kaganovich}%
\email{guendel@bgumail.bgu.ac.il , alexk@bgumail.bgu.ac.il}
\affiliation{%
Department of Physics, Ben-Gurion University of the Negev \\
P.O.Box 653, IL-84105 ~Beer-Sheva, Israel
}%

\author{E. Nissimov and S. Pacheva}%
\email{nissimov@inrne.bas.bg , svetlana@inrne.bas.bg}
\affiliation{%
Institute for Nuclear Research and Nuclear Energy,
Bulgarian Academy of Sciences \\
Boul. Tsarigradsko Chausee 72, BG-1784 ~Sofia, Bulgaria
}%

\begin{abstract}
We introduce and study in some detail the properties of a novel class of 
Weyl-conformally invariant $p$-brane theories which describe intrinsically 
lightlike branes for any odd world-volume dimension. 
Their dynamics significantly differs from that of the ordinary (conformally 
non-invariant) Nambu-Goto $p$-branes. We present
explicit solutions of the Weyl-invariant lightlike brane- (\textsl{WILL}-brane)
equations of motion in various gravitational models of physical
relevance exhibiting various new phenomena. In $D\! =\! 4$ the
\textsl{WILL}-membrane serves as a material and charged source for gravity and
electromagnetism in the coupled Einstein-Maxwell-WILL-membrane system; it
automatically positions itself on (``straddles'') the common event horizon of
the corresponding matching black hole solutions, thus providing an explicit
dynamical realization of the membrane paradigm in black hole physics.
In product spaces of interest in Kaluza-Klein theories the
\textsl{WILL}-brane wraps non-trivially around the compact (internal)
dimensions and still describes massless mode dynamics in the non-compact
(space-time) dimensions. Due to nontrivial variable size of the internal compact 
dimensions we find new types of physically interesting solutions describing
massless brane modes trapped on bounded planar circular orbits with non-trivial angular 
momentum, and with linear dependence between energy and angular momentum.
\end{abstract}

\pacs{11.25.-w, 04.70.-s, 04.50.+h}

\maketitle

\section{\label{sec:intro}Introduction}

In the recent years there has been a considerable interest in the study of
higher-dimensional extended objects motivated by various developments in
string theory, gravity, astrophysics and cosmology. 

In non-perturbative string theory there arise several types of higher-dimensional 
membranes ($p$-branes, $Dp$-branes) which play a crucial role in the description of
string dualities, microscopic physics of black holes, gauge 
theory/gravity correspondence, large-radius compactifications of extra dimensions, 
cosmological brane-world scenarios in high-energy particle phenomenology, 
\textit{etc.} (for a background on string and brane theories, see 
refs.\ct{brane-string-rev}). 

In the context of black hole physics, the so called ``membrane paradigm'' 
\ct{membrane-paradigm} appears to be a quite effective treatment of the physics of 
a black hole horizon. Furthermore, the thin-wall description of domain walls coupled 
to gravity \ct{Israel-66,Barrabes-Israel-Hooft} is able to provide neat
models for many cosmological and astrophysical effects.

It seems therefore of fundamental importance that all kinds of higher-dimensional 
extended objects, which could be consistently formulated, and their possible role 
in the various areas of physics should be thoroughly investigated.

Lightlike membranes are indeed of great importance in general relativity and have been
extensively studied from a phenomenological point of view 
\ct{Israel-66,Barrabes-Israel-Hooft}, \textsl{i.e.}, by introducing them
without specifying the Lagrangian dynamics from which they may
originate. These lightlike membranes have been treated as a source of gravity
enabling the formulation of important effects in the context of black hole
physics.

In the present paper we develop in some detail a new field-theoretic approach for 
a systematic description of the dynamics of lightlike $p$-branes starting from
a concise Weyl-conformally invariant action. Part of the results have been previously
reported in shorter form in refs.\ct{will-brane-kiten-zlatibor}. Our
approach is based on the general idea of employing alternative non-Riemannian
integration measures (volume-forms) in the actions of generally-covariant
(reparametrization-invariant) field theories instead of (or, more generally,
on equal footing with) the standard Riemannian volume forms. This idea has
been first proposed and applied in the context of four-dimensional theories
involving gravity \ct{TMT-basic} by introducing a new class of ``two-measure''
gravitational models. It has been demonstrated that the latter models are
capable to provide plausible solutions for a broad array of basic problems
in cosmology and particle physics, such as:
(i) scale invariance and its dynamical breakdown; (ii) spontaneous generation of
dimensionful fundamental scales; (iii) the cosmological constant problem;  
(iv) the problem of fermionic families; (v) applications to dark energy problem and
modern cosmological brane-world scenarios.
For a detailed discussion we refer to the series of papers \ct{TMT-basic,TMT-recent}.

Subsequently, the idea of employing an alternative non-Riemannian integration
measure was applied systematically to string, $p$-brane and $Dp$-brane models
\ct{m-string}. The main feature of these new classes of modified
string/brane theories is the appearance of the pertinent string/brane
tension as an additional dynamical degree of freedom beyond the usual string/brane
physical degrees of freedom, instead of being introduced \textsl{ad hoc} as
a dimensionful scale. The dynamical string/brane tension acquires the
physical meaning of a world-sheet electric field strength (in the string
case) or world-volume $(p+1)$-form field strength (in the $p$-brane case) and
obeys Maxwell (Yang-Mills) equations of motion or their higher-rank
anti\-symmetric tensor gauge field analogues, respectively. As a result of the
latter property the modified-measure string model with dynamical tension
yields a simple classical mechanism of ``color'' charge confinement \ct{m-string}. 

One drawback of modified-measure $p$-brane and $Dp$-brane models,
similarly to a drawback of ordinary Nambu-Goto $p$-branes, is that Weyl-conformal
invariance is lost beyond the simplest string case ($p\! =\! 1$). 
On the other hand, it turns out that the form of the action of the modified-measure 
string model with dynamical tension suggests a natural way to construct explicitly a 
substantially new class of {\em Weyl-conformally invariant} $p$-brane models 
{\em for any} $p$ \ct{will-brane-kiten-zlatibor}. The most profound property of the 
latter models is that for any even $p$ they describe the dynamics of inherently 
{\em lightlike} $p$-branes which makes them significantly different
both from the standard Nambu-Goto (or Dirac-Born-Infeld) branes as well as from 
their modified versions with dynamical string/brane tensions \ct{m-string} 
mentioned above.

Let us note that various papers have previously appeared in the literature 
\ct{Weyl-reform} where the standard Weyl-conformally non-invariant Nambu-Goto $p$-brane
action (Eq.\rf{stand-brane-action} below) and its supersymmetric counterparts were 
reformulated in a formally Weyl-invariant form by means of introducing auxiliary
non-dynamical fields with a non-trivial transformation properties under
Weyl-conformal symmetry appropriately tuned up to compensate for the Weyl
non-invariance with respect to the original dynamical degrees of freedom.
However, one immediately observes that the latter formally Weyl-invariant
$p$-brane actions {\em do not change} the dynamical content of the standard
Nambu-Goto $p$-branes (describing inherently {\em massive} modes). This is in sharp 
contrast to the new Weyl-conformally invariant $p$-brane models introduced and
studied in detail below, which describe {\em intrinsically lightlike} 
$p$-branes for any even $p$. In what follows we will use for the latter the
acronym \textsl{WILL-branes} (Weyl-invariant lightlike branes).

In the present paper we will demonstrate that \textsl{WILL}-branes can play
a very interesting role in diverse areas of physics. 
We begin with a short review of the
concept of alternative non-Riemannian volume form (integration measure) in
the context of string and $p$-brane models (Section \ref{sec:m-string}).
In Section \ref{sec:WI-action}, after a brief reminder of the standard
Polyakov-type formulation of ordinary Nambu-Goto $p$-branes, we introduce
and describe the Lagrangian formulation of the new class of inherently
Weyl-invariant $p$-branes for any $p$ and exhibit their intrinsic lightlike
nature when $p$ is even (\textsl{WILL}-branes). In Section \ref{sec:WILL-membrane}
and forward we study in detail the properties of \textsl{WILL}-membranes
(\textsl{i.e.}, for $p=2$), in particular, we introduce a natural coupling
of the \textsl{WILL}-membrane to external space-time electromagnetic fields.

The role of \textsl{WILL}-membranes in the context of gravity is discussed in
Sections \ref{sec:WILL-solutions} and \ref{sec:EM-WILL}. When moving as a test brane 
in a $D\! =\! 4$ black hole gravitational background  the \textsl{WILL}-membrane 
($p\! =\! 2$) automatically locates itself on the event horizon 
(Section \ref{sec:WILL-solutions}). Furthermore, as shown in Section \ref{sec:EM-WILL},
the \textsl{WILL}-membrane can serve as a material and charged 
source for gravity and electromagnetism in the coupled Einstein-Maxwell-WILL-membrane
system. We derive a self-consistent solution where the \textsl{WILL}-membrane locates
itself on (``straddles'') the common event horizon of two black holes 
(Reissner-Nordstr\"{o}m in the exterior and Schwarzschild in the interior).
Therefore, the \textsl{WILL}-membrane provides an explicit dynamical realization of
the membrane paradigm in black hole physics \ct{membrane-paradigm}.

The role of \textsl{WILL}-membranes in the context of Kaluza-Klein theories is 
studied in Section \ref{sec:KK} where we consider \textsl{WILL}-membrane
dynamics in higher-dimensional product-type space-time. It is shown that 
the \textsl{WILL}-membrane describes massless particle-like modes while acquiring 
non-trivial Kaluza-Klein quantum numbers. When the size of extra compact
dimensions is constant the motion of these massless brane modes is
indistinguishable from that of ordinary massless point-particles with respect to the
projected $D\! =\!4$ world. An interesting new feature arises
when the size of the extra compact dimensions has non-trivial space dependence.
In this case we find an explicit solution describing massless 
particle-like brane mode motion on the non-compact $D\! =\! 4$ space-time,
where the modes are trapped on bounded planar circular orbits with a linear 
relation between energy and angular momentum, while winding non-trivially the 
extra compact dimensions. The latter feature is inaccessible in standard 
Kaluza-Klein models.

The last Section collects some conclusions and outlook for future studies of the 
role and further aspects of \textsl{WILL}-brane dynamics. 

\section{\label{sec:m-string}String and Brane Models with a Modified
World-Sheet/World-Volume Integration Measure}

The modified-measure bosonic string model is given by the following action 
\ct{m-string}:
\br S = - \int d^2\s\,\P (\vp)
\Bigl\lb \h\g^{ab} \pa_a X^{\m} \pa_b X^{\n}G_{\m\n}(X) -
\frac{\vareps^{ab}}{2\sqrt{-\g}} F_{ab}(A)\Bigr\rb
\nonu \\
+ \int d^2\s\,\sqrt{-\g} A_a J^a  \qquad; \quad
J^a = \frac{\vareps^{ab}}{\sqrt{-\g}}\pa_b u \; ,
\lab{m-string}
\er
with the notations:
\be
\P (\vp) \equiv \h \vareps_{ij} \vareps^{ab} \pa_a \vp^i \pa_b \vp^j
\quad ,\quad F_{ab} (A) = \pa_{a} A_{b} - \pa_{b} A_{a} \; .
\lab{m-string-notat}
\ee
Here $\vp^i$ denote auxiliary world-sheet scalar fields,
$\g_{ab}$ indicates the intrinsic Riemannian world-sheet metric with
$\g = \det\Vert\g_{ab}\Vert$~ and $G_{\m\n}(X)$ is the Riemannian metric of
the embedding space-time ($a,b=0,1; i,j=1,2; \m,\n =0,1,\ldots,D-1$).

In action \rf{m-string} we notice the following differences with respect to the standard 
Nambu-Goto string (in the Polyakov-like formulation) :
\begin{itemize}
\item
New non-Riemannian integration measure density $\P (\vp)$ instead of $\sqrt{-\g}$;
\item
Dynamical string tension $T \equiv \frac{\P (\vp)}{\sqrt{-\g}}$ instead of
\textsl{ad hoc} dimensionful constant;
\item
Auxiliary world-sheet gauge field $A_a$ in a would-be ``topological'' term
$\int d^2\s\, \frac{\P (\vp)}{\sqrt{-\g}} \h\vareps^{ab} F_{ab}(A)$;
\item
Optional natural coupling of auxiliary $A_a$ to external conserved world-sheet
electric current $J^a$ (see last equality in \rf{m-string} and
Eq.\rf{Maxwell-like-eqs} below).
\end{itemize}

The modified string model \rf{m-string} is Weyl-conformally
invariant similarly to the standard Polyakov formulation.
Here Weyl-conformal symmetry is given by Weyl rescaling of
$\g_{ab}$ supplemented with a special diffeomorphism in
$\vp$-target space: \be \g_{ab} \longrightarrow \g^{\pr}_{ab} =
\rho\,\g_{ab}  \quad ,\quad \vp^{i} \longrightarrow \vp^{\pr\, i}
= \vp^{\pr\, i} (\vp) \;\; \mathrm{with} \;\; \det \Bigl\Vert
\frac{\pa\vp^{\pr\, i}}{\pa\vp^j} \Bigr\Vert = \rho \; .
\lab{Weyl-conf} \ee

The dynamical string tension appears as a canonically conjugated momentum
with respect to $A_1$:
$\pi_{A_1} \equiv \partder{\cL}{\dot{A_1}} = \frac{\P (\vp)}{\sqrt{-\g}}
\equiv T$, \textsl{i.e.}, $T$ has the meaning of a
\textit{world-sheet electric field strength}, and the equations of motion with respect to
auxiliary gauge field $A_a$ look exactly as $D=2$ Maxwell eqs.:
\be
\frac{\vareps^{ab}}{\sqrt{-\g}}\pa_b T + J^a = 0 \; .
\lab{Maxwell-like-eqs}
\ee
In particular, for $J^a= 0$ :
\be
\vareps^{ab} \pa_{b} \Bigl(\frac{\P (\vp)}{\sqrt{-\g}}\Bigr) = 0 \qquad,\quad
\frac {\P (\vp)}{\sqrt{-\gamma}}  \equiv T = \textrm{const.}\; ,
\lab{Maxwell-like-eqs-0}
\ee
one gets a {\em spontaneously induced} constant string tension.
Furthermore, when the modified string couples to point-like charges on the
world-sheet (\textsl{i.e.}, $J^0 {\sqrt{-\gamma}} = \sum_i e_i \d (\s - \s_i)$
in \rf{Maxwell-like-eqs}) one obtains classical charge {\em confinement}:
$\sum_i e_i = 0$.

The above charge confinement mechanism has also been generalized
in \ct{m-string} to the case of coupling the modified string model
(with dynamical tension) to non-Abelian world-sheet ``color''
charges. The latter is achieved as follows. Notice the following
identity in 2D involving Abelian gauge field $A_a$: 
\be
\frac{\vareps^{ab}}{2\sqrt{-\g}} F_{ab}(A) = 
\sqrt{-\h F_{ab}(A) F_{cd}(A) \g^{ac}\g^{bd}} \; . 
\lab{2D-id} 
\ee 
Using \rf{2D-id} the extension of the action \rf{m-string} to the non-Abelian case is
straightforward: 
\be 
S = - \int d^2\s \,\P (\vp) \Bigl\lb \h \g^{ab} \pa_a X^{\m} \pa_b X^\n G_{\m\n}(X) - 
\sqrt{-\h \Tr (F_{ab}(A)F_{cd}(A)) \g^{ac}\g^{bd}}\Bigr\rb + 
\int d^2\s\,\Tr \( A_a j^a\) \; ,
\lab{m-string-NA} 
\ee 
with $F_{ab}(A) = \pa_a A_b - \pa_b A_c + i \bigl\lb A_a,\, A_b\bigr\rb$.
The model \rf{m-string-NA} shares the same principal properties as the model
\rf{m-string} -- dynamical generation of string tension as an additional degree
of freedom, non-Abelian ``color'' charge confinement already on the classical 
level, \textsl{etc} .

Similar construction has also been proposed in \ct{m-string}
for higher-dimensional modified-measure $p$- and $Dp$-brane models
whose brane tension appears as an additional dynamical degree of
freedom. On the other hand, like the standard Nambu-Goto branes,
they are Weyl-conformally {\em non}-invariant and describe
dynamics of {\em massive} modes.

\section{\label{sec:WI-action}Weyl-Invariant Branes}

\subsection{Standard Nambu-Goto Branes}

Before proceeding to the main exposition,
let us briefly recall the standard Polyakov-type formulation of the ordinary
(bosonic) Nambu-Goto $p$-brane action:
\be
S = -{T\o 2}\int d^{p+1}\s\,\sqrt{-\g}\, \Bigl\lb
\g^{ab}\pa_a X^\m \pa_b X^\n G_{\m\n}(X) - \L (p-1)\Bigr\rb \; .
\lab{stand-brane-action}
\ee
Here $\g_{ab}$ is the ordinary Riemannian metric on the $p+1$-dimensional
brane world-volume with $\g \equiv \det\v\v \g_{ab}\v\v$. The world-volume indices
$a,b=0,1,\ldots ,p$ ;
~$G_{\m\n}$ denotes the Riemannian metric in the embedding space-time with
space-time indices $\m,\n=0,1,\ldots ,D-1$.
$T$ is the given \textsl{ad hoc} brane tension; the constant
$\L$ can be absorbed by rescaling $T$ (see below Eq.\rf{stand-brane-action-NG}).
The equations of motion with respect to $\g^{ab}$ and $X^\m$ read:
\be
T_{ab} \equiv \( \pa_a X^\m \pa_b X^\n -
\h \g_{ab} \g^{cd}\pa_c X^\m \pa_d X^\n \) G_{\m\n} + \g_{ab} {\L\o 2}(p-1)
= 0 \; ,
\lab{stand-brane-gamma-eqs}
\ee
\be
\pa_a \(\sqrt{-\g}\g^{ab}\pa_b X^\m\) +
\sqrt{-\g}\g^{ab}\pa_a X^\n \pa_b X^\l \G^\m_{\n\l} = 0  \; ,
\lab{stand-brane-X-eqs}
\ee
where:
\be
\G^\m_{\n\l}=\h G^{\m\k}\(\pa_\n G_{\k\l}+\pa_\l G_{\k\n}-\pa_\k G_{\n\l}\)
\lab{affine-conn}
\ee
is the Christoffel connection for the external metric.

Eqs.\rf{stand-brane-gamma-eqs} when $p \neq 1$ imply:
\be
\L \g_{ab} = \pa_a X^\m \pa_b X^\n G_{\m\n} \; ,
\lab{stand-brane-metric-eqs}
\ee
which in turn allows to rewrite Eq.\rf{stand-brane-gamma-eqs} as:
\be
T_{ab} \equiv \( \pa_a X^\m \pa_b X^\n -
{1\o {p+1}} \g_{ab} \g^{cd}\pa_c X^\m \pa_d X^\n \) G_{\m\n} = 0 \; .
\lab{stand-brane-gamma-eqs-Pol}
\ee
Furthermore, using \rf{stand-brane-metric-eqs} the Polyakov-type brane action
\rf{stand-brane-action} becomes on-shell equivalent to the Nambu-Goto-type brane
action:
\be
S = - T \L^{-{{p-1}\o 2}} \int d^{p+1}\s\,
\sqrt{-\det\v\v \pa_a X^\m \pa_b X^\n G_{\m\n} \v\v} \; .
\lab{stand-brane-action-NG}
\ee

Let us note the following properties of standard Nambu-Goto $p$-branes
manifesting their crucial differences with respect to the Weyl-conformally invariant
branes discussed below.  Eq.\rf{stand-brane-metric-eqs} tells us that: (i) the
induced metric on the Nambu-Goto $p$-brane world-volume is {\em non-singular};
(ii) standard Nambu-Goto $p$-branes describe intrinsically {\em massive} modes.

\subsection{Weyl-Invariant Branes: Action and Equations of Motion}
Identity \rf{2D-id} and the modified-measure string action \rf{m-string-NA} 
naturally suggest how to construct \textbf{Weyl-invariant} $p$-brane models for 
any $p$. Namely, we consider the following novel class of $p$-brane actions:
\be
S = - \int d^{p+1}\s \,\P (\vp)
\Bigl\lb \h \g^{ab} \pa_a X^{\m} \pa_b X^{\n} G_{\m\n}(X)
- \sqrt{F_{ab}(A) F_{cd}(A) \g^{ac}\g^{bd}}\Bigr\rb
\lab{WI-brane}
\ee
\be
\P (\vp) \equiv \frac{1}{(p+1)!} \vareps_{i_1\ldots i_{p+1}}
\vareps^{a_1\ldots a_{p+1}} \pa_{a_1} \vp^{i_1}\ldots \pa_{a_{p+1}} \vp^{i_{p+1}}
\; ,
\lab{mod-measure-p}
\ee
where notations similar to those in \rf{m-string} are used
(here $a,b=0,1,\ldots,p; i,j=1,\ldots,p+1$). In particular, $\vp^i$ are
world-volume scalar fields which are the building blocks of the
non-Riemannian integration measure \rf{mod-measure-p}.

The above action \rf{WI-brane} is invariant under Weyl-conformal symmetry
(the same as in the dynamical-tension string model \rf{m-string}):
\be
\g_{ab} \longrightarrow \g^{\pr}_{ab} = \rho\,\g_{ab}  \quad ,\quad
\vp^{i} \longrightarrow \vp^{\pr\, i} = \vp^{\pr\, i} (\vp)
\;\; \mathrm{with} \;\;
\det \Bigl\Vert \frac{\pa\vp^{\pr\, i}}{\pa\vp^j} \Bigr\Vert = \rho \; .
\lab{Weyl-conf-p}
\ee

Let us note the following significant differences of \rf{WI-brane} with respect to the standard
Nambu-Goto $p$-branes (in the Polyakov-like formulation) :
\begin{itemize}
\item
New non-Riemannian integration measure density $\P (\vp)$ instead of $\sqrt{-\g}$,
and {\em no}~ ``cosmological-constant'' term ($(p-1)\sqrt{-\g}$);
\item
Variable brane tension $\chi \equiv \frac{\P (\vp)}{\sqrt{-\g}}$
which is now Weyl-conformally {\em gauge dependent}: $ \chi \to \rho^{\h(1-p)}\chi$;
\item
Auxiliary world-volume gauge field $A_a$ in a ``square-root''
Maxwell-type term \footnote{``Square-root'' Maxwell (Yang-Mills) action in
$D=4$ was originally introduced in the first ref.\ct{Spallucci}
and later generalized to ``square-root'' actions of higher-rank
antisymmetric tensor gauge fields in $D\geq 4$ in the second and
third refs.\ct{Spallucci}.}; the latter can be straightforwardly
generalized to the non-Abelian case -- $\sqrt{-\Tr \( F_{ab}(A)
F_{cd}(A)\) \g^{ac}\g^{bd}}$ similarly to \rf{m-string-NA};
\item
Natural optional couplings of the auxiliary gauge field $A_a$ to external
world-volume ``color'' charge currents $j^a$ as in \rf{m-string-NA};
\item
The action \rf{WI-brane} is manifestly Weyl-conformal invariant for {\em any} $p$;
it describes {\em intrinsically lightlike} $p$-branes for any even $p$, as
it will be shown below.
\end{itemize}

In what follows we shall frequently use the short-hand notations:
\be
\(\pa_a X \pa_b X\) \equiv \pa_a X^\m \pa_b X^\n G_{\m\n}\quad ,\quad
\sqrt{FF\g\g} \equiv \sqrt{F_{ab} F_{cd} \g^{ac}\g^{bd}} \; .
\lab{short-hand}
\ee
Employing \rf{short-hand} the equations of motion with respect to measure-building
auxiliary scalars $\vp^i$ and with respect to $\g^{ab}$ read, respectively:
\be
\h \g^{cd}\(\pa_c X \pa_d X\) - \sqrt{FF\g\g} = M \; \Bigl( = \mathrm{const.}\Bigr)
\; ,
\lab{phi-eqs}
\ee
\be
\h\(\pa_a X \pa_b X\) + \frac{F_{ac}\g^{cd} F_{db}}{\sqrt{FF\g\g}} = 0 \; .
\lab{gamma-eqs}
\ee
Taking the trace in \rf{gamma-eqs} implies $M=0$ in Eq.\rf{phi-eqs}.

Next we have the following equations of motion with respect to auxiliary gauge field $A_a$
and with respect to $X^\m$, respectively:
\be
\pa_b \(\frac{F_{cd}\g^{ac}\g^{bd}}{\sqrt{FF\g\g}} \P (\vp)\) = 0 \; ,
\lab{A-eqs}
\ee
\be
\pa_a \(\P (\vp) \g^{ab}\pa_b X^\m\) +
\P (\vp) \g^{ab}\pa_a X^\n \pa_b X^\l \G^\m_{\n\l} = 0 \; ,
\lab{X-eqs}
\ee
where $\G^\m_{\n\l}$ is the Christoffel  connection corresponding to the external
space-time metric $G_{\m\n}$ as in \rf{affine-conn}.

\subsection{\label{sec:WILL-branes}Intrinsically Lightlike Branes}
Let us consider the $\g^{ab}$-equations of motion \rf{gamma-eqs}.
$F_{ab}$ is an anti-symmetric $(p+1)\times (p+1)$ matrix,
therefore, $F_{ab}$ is {\em not invertible} in any odd $(p+1)$ --
it has at least one zero-eigenvalue vector $V^a$ ($F_{ab}V^b = 0$). Therefore, for 
any odd $(p+1)$ the induced metric:
\be 
g_{ab} \equiv \(\pa_a X \pa_b X\) \equiv \pa_a X^\m \pa_b X^\n G_{\m\n}(X) 
\lab{ind-metric} 
\ee 
on the world-volume of the Weyl-invariant brane \rf{WI-brane} is {\em singular} as 
{\em opposed} to the ordinary Nambu-Goto brane (where the induced
metric is proportional to the intrinsic Riemannian world-volume
metric, cf. Eq.\rf{stand-brane-metric-eqs}): 
\be 
\(\pa_a X \pa_b X\) V^b = 0 \quad ,\quad \mathrm{i.e.}\;\; \(\pa_V X \pa_V X\) = 0
\;\; ,\;\; \(\pa_{\perp} X \pa_V X\) = 0 \; , 
\lab{LL-constraints}
\ee 
where $\pa_V \equiv V^a \pa_a$ and $\pa_{\perp}$ are derivatives along the tangent 
vectors in the complement of the tangent vector field $V^a$.

Thus, we arrive at the following important conclusion:
every point on the world-surface of the Weyl-invariant $p$-brane \rf{WI-brane}
(for odd $(p+1)$) moves with the speed of light in a time-evolution along the
zero-eigenvalue vector-field $V^a$ of $F_{ab}$. Therefore, we will name
\rf{WI-brane} (for odd $(p+1)$) by the acronym {\em WILL-brane}
(Weyl-Invariant Lightlike-brane) model.

\subsection{Dual Formulation of {\em WILL}-Branes}
The $A_a$-equations of motion \rf{A-eqs} can be solved in terms of
$(p-2)$-form gauge potentials $\L_{a_1\ldots a_{p-2}}$ dual with respect to
$A_a$. The respective field-strengths are related as follows:
\br 
F_{ab}(A)= -\frac{1}{\chi}\,\frac{\sqrt{-\g}\,\vareps_{abc_1\ldots
c_{p-1}}}{2(p-1)} \g^{c_1 d_1}\ldots \g^{c_{p-1} d_{p-1}} \,
F_{d_1\ldots d_{p-1}}(\L) \,\g^{cd} \(\pa_c X \pa_d X\) \; ,
\lab{dual-strength-rel} \\
\chi^2 = - \frac{2}{(p-1)^2}\, \g^{a_1 b_1}\ldots \g^{a_{p-1} b_{p-1}}
F_{a_1\ldots a_{p-1}}(\L) F_{b_1\ldots b_{p-1}}(\L) \; ,
\lab{chi-2}
\er
where $\chi \equiv \frac{\P (\vp)}{\sqrt{-\g}}$ is the variable brane tension, and:
\be
F_{a_1\ldots a_{p-1}}(\L) = (p-1) \pa_{[a_1} \L_{a_2\ldots a_{p-1}]}
\lab{dual-strength}
\ee
is the $(p-1)$-form dual field-strength.

All equations of motion \rf{phi-eqs}--\rf{X-eqs} can be equivalently derived from 
the following {\em dual} {\em WILL}-brane action:
\be
S_{\mathrm{dual}} \lb X,\g,\L\rb = - \h \int d^{p+1}\s\, \chi (\g,\L) \sqrt{-\g}
\g^{ab}\pa_a X^\m \pa_b X^\n G_{\m\n}(X)
\lab{WI-brane-dual}
\ee
with $\chi (\g,\L)$ given in \rf{chi-2} above. In particular, in terms of
the dual gauge fields \rf{dual-strength} Eqs.\rf{gamma-eqs}-\rf{A-eqs} read:
\be
\(\pa_a X \pa_b X\) + \h \g^{cd}\(\pa_c X \pa_d X\)\, \biggl\lbrack - \g_{ab} +
(p-1) \frac{\g^{a_1 b_1}\ldots\g^{a_{p-2} b_{p-2}} F_{a a_1\ldots a_{p-2}}(\L)
F_{b b_1\ldots b_{p-2}}(\L)}{\g^{c_1 d_1}\ldots\g^{c_{p-1} d_{p-1}} 
F_{c_1\ldots c_{p-1}}(\L) F_{d_1\ldots d_{p-1}}(\L)} \biggr\rbrack = 0  \; ,
\lab{gamma-dual-eqs}
\ee
\be
\pa_b \biggl( \g^{a_1 b_1}\ldots\g^{a_{p-2} b_{p-2}} \g^{b b_{p-1}}
F_{b_1\ldots b_{p-1}}(\L) \frac{1}{\chi (\g,\L)} \sqrt{-\g} \g^{cd}\(\pa_c X \pa_d X\)
\biggr) = 0 \; ,
\lab{A-dual-eqs}
\ee
with $\chi (\g,\L)$ as in \rf{chi-2}.

\section{\label{sec:WILL-membrane}Special case $p=2$. Coupling to External
Electromagnetic Field}

\subsection{{\em WILL}-Membrane}

Henceforth we will explicitly consider the special case $p=2$ of \rf{WI-brane},
\textsl{i.e.}, the Weyl-invariant lightlike membrane model:
\be
S = - \int d^3\s \,\P (\vp)
\Bigl\lb \h \g^{ab} \pa_a X^{\m} \pa_b X^{\n} G_{\m\n}(X)
- \sqrt{F_{ab}(A) F_{cd}(A) \g^{ac}\g^{bd}}\Bigr\rb
\lab{WILL-membrane-0}
\ee
\be
\P (\vp) \equiv \frac{1}{3!} \vareps_{ijk}
\vareps^{abc} \pa_a \vp^i \pa_b \vp^j \pa_c \vp^k \quad ,\quad 
a,b,c =0,1,2\; ,\; i,j,k=1,2,3  \; .
\lab{mod-measure-3}
\ee
The associated {\em WILL}-membrane dual action (particular case of \rf{WI-brane-dual}
for $p=2$) reads:
\be
S_{\mathrm{dual}} = - \h \int d^3\s\, \chi (\g,u)\,\sqrt{-\g}
\g^{ab} \pa_a X^{\m} \pa_b X^{\n} G_{\m\n}(X)   \qquad ,\quad
\chi (\g,u) \equiv \sqrt{-2\g^{cd}\pa_c u \pa_d u} \; ,
\lab{WILL-membrane}
\ee
where $u$ is the dual ``gauge'' potential with respect to $A_a$:
\be
F_{ab}(A) = - \frac{1}{2\chi (\g,u)} \sqrt{-\g} \vareps_{abc} \g^{cd}\pa_d u\,
\g^{ef}\!\(\pa_e X \pa_f X\)  \; .
\lab{dual-strenght-rel-3}
\ee
$S_{\mathrm{dual}}$ is manifestly Weyl-invariant (under $\g_{ab} \to \rho\g_{ab}$).

The equations of motion with respect to $\g^{ab}$, $u$ (or $A_a$), and
$X^\m$ read accordingly (using again short-hand notation \rf{ind-metric}) :
\be 
\(\pa_a X \pa_b X\) + \h \g^{cd}\(\pa_c X \pa_d X\) 
\(\frac{\pa_a u \pa_b u }{\g^{ef} \pa_e u \pa_f u} - \g_{ab}\) = 0 \; ,
\lab{gamma-eqs-3} 
\ee 
\be 
\pa_a \(\,\frac{\sqrt{-\g}\g^{ab}\pa_b u}{\chi (\g,u)}\,\g^{cd}\(\pa_c X \pa_d X\)\,\)
= 0 \; , 
\lab{u-eqs} 
\ee 
\be 
\pa_a \(\chi (\g,u)\,\sqrt{-\g} \g^{ab}\pa_b X^\m \) + 
\chi (\g,u)\,\sqrt{-\g} \g^{ab}\pa_a X^\n \pa_b X^\l \G^\m_{\n\l} = 0 \; . 
\lab{X-eqs-3}
\ee
Eq.\rf{gamma-eqs-3} represents the relation between the intrinsic metric
$\g_{ab}$ and the induced metric $\(\pa_a X \pa_b X\)$. However, the last factor 
in brackets on the l.h.s. of \rf{gamma-eqs-3} is a projector implying that the 
induced metric $g_{ab} \equiv \(\pa_a X \pa_b X\)$ has zero-mode eigenvector 
$V^a =\g^{ab}\pa_b u$.

\subsection{Gauge-Fixed Constraints and Equations of Motion}
The invariance under world-volume reparametrizations allows to introduce the
following standard (synchronous) gauge-fixing conditions:
\be
\g^{0i} = 0 \;\; (i=1,2) \quad ,\quad \g^{00} = -1 \; .
\lab{gauge-fix}
\ee
Using \rf{gauge-fix} we can easily find solutions of Eq.\rf{u-eqs} for the
dual ``gauge potential'' $u$ in spite of its high non-linearity by taking the
following ansatz:
\be
u (\t,\s^1,\s^2) = \frac{T_0}{\sqrt{2}}\t  \; ,
\lab{u-ansatz}
\ee
Here $T_0$ is an arbitrary integration constant with the dimension of membrane
tension. In particular:
\be
\chi \equiv \sqrt{-2\g^{ab}\pa_a u \pa_b u} = T_0
\lab{chi-0}
\ee
The ansatz \rf{u-ansatz} means that we take $\t\equiv\s^0$ to be evolution
parameter along the zero-eigenvalue vector-field of the induced metric on the brane
($V^a = \g^{ab}\pa_b u = \mathrm{const.}\,(1,0,0)$). Also, in terms of the
original gauge field $A_a$ (cf. relation \rf{dual-strenght-rel-3}) Eq.\rf{u-ansatz} 
implies vanishing of the world-volume ``electric'' field-strength $F_{0i}(A) = 0$.

The ansatz for $u$ \rf{u-ansatz} together with the gauge choice for $\g_{ab}$
\rf{gauge-fix} brings the equations of motion with respect to $\g^{ab}$, $u$ (or $A_a$) and
$X^\m$ in the following form
(recall $\(\pa_a X \pa_b X\) \equiv \pa_a X^\m \pa_b X^\n G_{\m\n}$):
\be
\(\pa_0 X \pa_0 X\) = 0 \quad ,\quad \(\pa_0 X \pa_i X\) = 0  \; ,
\lab{constr-0}
\ee
\be
\(\pa_i X\pa_j X\) - \h \g_{ij} \g^{kl}\(\pa_k X\pa_l X\) = 0  \; ,
\lab{constr-vir}
\ee
(notice that Eqs.\rf{constr-vir} look exactly like the classical (Virasoro)
constraints for an Euclidean string theory with world-sheet parameters $(\s^1,\s^2)$);
\be
\pa_0 \(\sqrt{\g_{(2)}} \g^{kl}\(\pa_k X\pa_l X\)\) = 0  \; ,
\lab{u-eqs-fix}
\ee
where $\g_{(2)} = \det\Vert \g_{ij}\Vert$; 
\be
\Box^{(3)} X^\m + \( - \pa_0 X^\n \pa_0 X^\l +
\g^{kl} \pa_k X^\n \pa_l X^\l \) \G^{\m}_{\n\l} = 0  \; ,
\lab{X-eqs-3-fix}
\ee
where:
\be
\Box^{(3)} \equiv
- \frac{1}{\sqrt{\g^{(2)}}} \pa_0 \(\sqrt{\g^{(2)}} \pa_0 \) +
\frac{1}{\sqrt{\g^{(2)}}}\pa_i \(\sqrt{\g^{(2)}} \g^{ij} \pa_j \)   \; .
\lab{box-3}
\ee

Let us note that Eq.\rf{u-eqs-fix} is the only remnant from the $A_a$-equations of
motion \rf{A-eqs} and, in fact, it can easily be shown that \rf{u-eqs-fix} is
a consequence of the gauge-fixed constraints \rf{constr-0}-\rf{constr-vir} and
equations of motion \rf{X-eqs-3-fix}.

\subsection{Coupling to External Electromagnetic Field}
We can also extend the {\em WILL}-brane model \rf{WI-brane} via a
coupling to external space-time electromagnetic field $\cA_\m$.
The natural Weyl-conformal invariant candidate action reads (for $p=2$): 
\be 
S_{\mathrm{WILL-brane}} = - \int d^3\s \,\P (\vp)
\Bigl\lb \h \g^{ab} \pa_a X^{\m} \pa_b X^{\n} G_{\m\n} -
\sqrt{F_{ab} F_{cd} \g^{ac}\g^{bd}}\Bigr\rb - q\int d^3\s \,
\vareps^{abc} \cA_\m \pa_a X^\m F_{bc} \; . 
\lab{WILL-membrane+A}
\ee
The last Chern-Simmons-like term is a special case of a class
of Chern-Simmons-like couplings of extended objects to external
electromagnetic fields proposed in ref.\ct{Aaron-Eduardo}.

Instead of the action \rf{WILL-membrane+A} we can use its dual one (similar to
the simpler case Eq.\rf{WI-brane} versus Eq.\rf{WILL-membrane}):
\be
S^{\mathrm{dual}}_{\mathrm{WILL-brane}}
= - \h \int d^3\s\, \chi (\g,u,\cA)\,\sqrt{-\g} \g^{ab}\(\pa_a X \pa_b X\)  \; ,
\lab{WILL-membrane+A-dual}
\ee
where the variable brane tension $\chi \equiv \frac{\P (\vp)}{\sqrt{-\g}}$
is given by:
\be
\chi (\g,u,\cA) \equiv \sqrt{-2\g^{cd}\(\pa_c u - q \cA_c\)
\(\pa_d u - q \cA_d\)} \quad ,\;\; \cA_a \equiv \cA_\m \pa_a X^\m \; .
\lab{tension+A}
\ee
Here $u$ is the dual ``gauge'' potential with respect to $A_a$ and the corresponding
field-strength and dual field-strength are related as
(cf. Eq.\rf{dual-strenght-rel-3}) :
\be
F_{ab}(A) = - \frac{1}{2\chi (\g,u,\cA)} \sqrt{-\g} \vareps_{abc} \g^{cd}
\(\pa_d u - q\cA_d \)\,\g^{ef}\!\(\pa_e X \pa_f X\)  \; .
\lab{dual-strenght-rel-3-A}
\ee

The corresponding equations of motion with respect to $\g^{ab}$, $u$ (or $A_a$), and $X^\m$
read accordingly:
\be
\(\pa_a X \pa_b X\) + \h \g^{cd}\(\pa_c X \pa_d X\)
\(\frac{\(\pa_a u - q\cA_a\)\(\pa_b u -q\cA_b\) }
{\g^{ef} \(\pa_e u - q\cA_e\)\(\pa_f u - q\cA_f\)} - \g_{ab}\) = 0  \; ;
\lab{gamma-eqs+A}
\ee
\be
\pa_a \(\,\frac{\sqrt{-\g}\g^{ab}\(\pa_b u - q\cA_b\)}{\chi (\g,u,\cA)}\,
\g^{cd}\(\pa_c X \pa_d X\)\,\) = 0  \; ;
\lab{u-eqs+A}
\ee
\br
\pa_a \(\chi (\g,u,\cA)\,\sqrt{-\g} \g^{ab}\pa_b X^\m \) +
\chi (\g,u,\cA)\,\sqrt{-\g} \g^{ab}\pa_a X^\n \pa_b X^\l \G^\m_{\n\l}
\nonu \\
- \; q \vareps^{abc} F_{bc} \pa_a X^\n
\(\pa_\l \cA_\n - \pa_\n \cA_\l\)\, G^{\l\m} = 0  \; . \phantom{aaaaaaaa}
\lab{X-eqs+A}
\er

\section{\label{sec:WILL-solutions}{\em WILL}-Membrane Solutions in Various 
Gravitational Backgrounds}

\subsection{Example: {\em WILL}-Membrane in a PP-Wave Background}

As a first non-trivial example let us consider
\textsl{WILL}-membrane dynamics in an external background
generalizing the plane-polarized gravitational wave
(\textsl{pp-wave}): 
\be 
(ds)^2 = - dx^{+} dx^{-} - F(x^{+},x^I)\, (dx^{+})^2 + h_{IJ}(x^K) dx^I dx^J \; , 
\lab{gen-pp-wave} 
\ee 
(for the ordinary pp-wave $h_{IJ}(x^K) = \d_{IJ}$), and let us employ
in \rf{constr-0}--\rf{box-3} the following natural ansatz for
$X^\m$ (here $\s^0 \equiv \t$; $I=1,\ldots,D-2$) :
\be 
X^{-} = \t \quad, \quad
X^{+}=X^{+}(\t,\s^1,\s^2) \quad, \quad X^I = X^I (\s^1,\s^2) \; .
\lab{ansatz-pp-wave} 
\ee 
The non-zero affine connection symbols for the generalized pp-wave metric 
\rf{gen-pp-wave} are:
$\G^{-}_{++}=\pa_{+}F$, $\G^{-}_{+I}=\pa_{I}F$, $\G^{I}_{++}=\h
h^{IJ}\pa_{J}F$, and $\G^{I}_{JK}$ -- the ordinary Christoffel
symbols for the metric $h_{IJ}$ in the transverse dimensions.

It is straightforward to show that the solution does not depend on the form of
the pp-wave front $F(x^{+},x^I)$ and reads:
\be
X^{+}=X^{+}_0 = \mathrm{const.} \quad ,\quad
\g_{ij} \; \mathrm{are}\; \t\!-\!\mathrm{independent}\; ;
\lab{pp-wave-sol-1}
\ee
\be
\(\pa_i X^I \pa_j X^J - \h \g_{ij} \g^{kl} \pa_k X^I \pa_l X^J\)\, h_{IJ} = 0
\lab{pp-wave-sol-2}
\ee
\be
\frac{1}{\sqrt{\g^{(2)}}} \pa_i \(\sqrt{\g^{(2)}} \g^{ij} \pa_j X^I \)
+ \g^{kl} \pa_k X^J \pa_l X^K \G^{I}_{JK} = 0
\lab{pp-wave-sol-3}
\ee
The latter two equations for the transverse brane coordinates describe a string
moving in the $(D-2)$-dimensional Euclidean-signature transverse space.

\subsection{Example: {\em WILL}-Membrane in Spherically-Symmetric
Backgrounds}

Let us consider general $SO(3)$-symmetric background in $D=4$ embedding space-time:
\be
(ds)^2 =  - A(z,t) (dt)^2 + B(z,t) (dz)^2 + C (z,t) \( (d\th)^2 + \sin^2\th (d\p)^2\).
\lab{backgr}
\ee

The usual ansatz:
\br
X^0 \equiv t = \t \quad ,\quad
X^1 \equiv z = z (\t,\s^1,\s^2) \quad ,\quad
X^2 \equiv \th = \s^1 \quad ,\quad X^3 \equiv \p = \s^2
\lab{so3-ansatz} \\
\g_{ij} = a (\t) \( (d\s^1)^2 + \sin^2(\s^1) (d\s^2)^2\)
\phantom{aaaaaaaaaaaaa}
\nonu
\er
yields:

(i) equations for $z (\t,\s^1,\s^2)$ :
\be
\frac{\pa z}{\pa \t} = \pm \sqrt{\frac{A}{B}}  \quad ,\quad
\frac{\pa z}{\pa \s^i} = 0
\;\; ;
\lab{z-eqs}
\ee

(ii) a restriction on the background itself (comes from the gauge-fixed equations of
motion for the dual gauge potential $u$ \rf{u-eqs-fix}) :
\be
\frac{d C}{d\t} \equiv \(\partder{C}{t} \pm
\sqrt{\frac{A}{B}}\, \partder{C}{z}\)\bgv_{t=\t,\; z=z(\t)} = 0   \;\; ;
\lab{C-eq}
\ee

(iii) an equation for the conformal factor $a(\t)$ of the internal membrane metric:
\be
\pa_\t a + \(\frac{\partder{}{t}\sqrt{AB} \pm \pa_z A}{\sqrt{AB}}
\bgv_{t=\t,\, z=z(\t)}\)\, a(\t)
- \frac{\partder{}{t}C}{A}\bgv_{t=\t,\, z=z(\t)} = 0 \;\; .
\lab{a-eq}
\ee
Eq.\rf{C-eq} tells that the (squared) sphere radius $R^2 \equiv C (z,t)$ must remain
constant along the {\em WILL}-brane trajectory.


\subsection{Example: {\em WILL}-Membrane in Schwarzschild and
Reissner-Nordstr\"{o}m Black Holes}

Let us apply the results of Subsection V.B for static
spherically-symmetric gravitational background in $D=4$: 
\be
(ds)^2 = - A(r)(dt)^2 + B(r)(dr)^2 + r^2 \lb (d\th)^2 + \sin^2
(\th)\,(d\p)^2\rb \; . 
\lab{spherical-symm-metric} 
\ee
Specifically we have: 
\be A(r) = B^{-1}(r) = 1 - \frac{2GM}{r}
\lab{schwarzschild} 
\ee 
for Schwarzschild black hole, 
\be 
A(r) =
B^{-1}(r) = 1 - \frac{2GM}{r} + \frac{Q^2}{r^2} \lab{R-N} \ee for
Reissner-Nordstr\"{o}m black hole, \be A(r) = B^{-1}(r) = 1 - \k
r^2 
\lab{AdS} 
\ee 
for (anti-) de Sitter space, \textsl{etc.}.

In the case of \rf{spherical-symm-metric} Eqs.\rf{z-eqs}--\rf{C-eq} reduce to:
\be
\partder{r}{\t} = \pm A(r) \quad ,\quad \partder{r}{\s^i} = 0 \quad ,\quad
\partder{r}{\t} = 0
\lab{sol-spherical-1}
\ee
yielding:
\be
r = r_0 \equiv \mathrm{const.} \;\;,\;\; \mathrm{where} \quad A(r_0)=0  \; .
\lab{sol-spherical-2}
\ee
Further, Eq.\rf{a-eq} implies for the intrinsic {\em WILL}-membrane metric:
\be
\Vert\g_{ij}\Vert = c_0 \, e^{\mp \t/r_0}\,\twomat{1}{0}{0}{\sin^2 (\s^1)} \; ,
\lab{sol-spherical-3}
\ee
where $c_0$ is an arbitrary integration constant.

From \rf{sol-spherical-2} we conclude that the {\em WILL}-membrane with spherical
topology (and with exponentially blowing-up/deflating radius with respect to internal
metric, see Eq.\rf{sol-spherical-3}) automatically ``sits'' on (``straddles'')
the event horizon of the pertinent black hole in $D=4$ embedding space-time. This
conforms with the well-known general property of closed lightlike hypersurfaces
in $D=4$ (\textsl{i.e.}, their section with the hyper-plane $t$=const.~ being a
compact 2-dimensional manifold) which automatically serve as horizons \ct{LL-vol2}.
On the other hand, let us stress that our \textsl{WILL}-membrane model
\rf{WILL-membrane} provides an explicit {\em dynamical} realization of event horizons.


\section{\label{sec:EM-WILL}Coupled Einstein-Maxwell-{\em WILL}-Membrane
System: {\em WILL}-Membrane as a Source for Gravity and Electromagnetism}

We can extend the results from the previous section to the case of
the self-consistent Einstein-Maxwell-{\em WILL}-membrane system,
\textsl{i.e.}, we will consider the {\em WILL}-membrane as a
dynamical material and electrically charged source for gravity and
electromagnetism. The relevant action reads: 
\be 
S = \int\!\! d^4 x\,\sqrt{-G}\,\llb \frac{R (G)}{16\pi G_N} - 
\frac{1}{4} \cF_{\m\n}(\cA) \cF_{\k\l}(\cA) G^{\m\k} G^{\n\l}\rrb +
S_{\mathrm{WILL-brane}}  \; , 
\lab{E-M-WILL} 
\ee 
where $\cF_{\m\n}(\cA) = \pa_\m \cA_\n - \pa_\n \cA_\m$ is the space-time 
electromagnetic field-strength, and $S_{\mathrm{WILL-brane}}$ indicates the 
\textsl{WILL}-membrane action coupled to the space-time gauge field $\cA_\m$ -- 
either \rf{WILL-membrane+A} or its dual \rf{WILL-membrane+A-dual}.

The equations of motion for the \textsl{WILL}-membrane subsystem are of the same
form as Eqs.\rf{gamma-eqs+A}--\rf{X-eqs+A}. The Einstein-Maxwell equations of motion
read:
\be
R_{\m\n} - \h G_{\m\n} R = 8\pi G_N \( T^{(EM)}_{\m\n} + T^{(brane)}_{\m\n}\)\; ,
\lab{Einstein-eqs}
\ee
\be
\pa_\n \(\sqrt{-G}G^{\m\k}G^{\n\l} \cF_{\k\l}\) + j^\m = 0 \; ,
\lab{Maxwell-eqs}
\ee
where:
\be
T^{(EM)}_{\m\n} \equiv \cF_{\m\k}\cF_{\n\l} G^{\k\l} - G_{\m\n}\frac{1}{4}
\cF_{\r\k}\cF_{\s\l} G^{\r\s}G^{\k\l} \; ,
\lab{T-EM}
\ee
\be
T^{(brane)}_{\m\n} \equiv - G_{\m\k}G_{\n\l}
\int\!\! d^3 \s\, \frac{\d^{(4)}\Bigl(x-X(\s)\Bigr)}{\sqrt{-G}}\,
\chi\,\sqrt{-\g} \g^{ab}\pa_a X^\k \pa_b X^\l \; ,
\lab{T-brane}
\ee
\be
j^\m \equiv q \int\!\! d^3 \s\,\d^{(4)}\Bigl(x-X(\s)\Bigr)
\vareps^{abc} F_{bc} \pa_a X^\m \; .
\lab{brane-EM-current}
\ee

We find the following self-consistent spherically symmetric stationary solution
for the coupled Einstein-Maxwell-{\em WILL}-membrane system \rf{E-M-WILL}.
For the Einstein subsystem we have a solution:
\be
(ds)^2 = - A(r)(dt)^2 + A^{-1}(r)\,(dr)^2 +
r^2 \lb (d\th)^2 + \sin^2 (\th)\,(d\p)^2\rb  \; ,
\lab{spherical-symm-metric-b}
\ee
consisting of two different black holes with a {\em common} event horizon:
\begin{itemize}
\item
Schwarzschild black hole inside the horizon:
\be
A(r)\equiv A_{-}(r) = 1 - \frac{2GM_1}{r}\;\; ,\quad \mathrm{for}\;\;
r < r_0 \equiv r_{\mathrm{horizon}}= 2GM_1 \; .
\lab{Schwarzschild-metric-in}
\ee
\item
Reissner-Norstr\"{o}m black hole outside the horizon:
\be
A(r)\equiv A_{+}(r) = 1 - \frac{2GM_2}{r} + \frac{GQ^2}{r^2}\;\; ,
\quad \mathrm{for}\;\; r > r_0 \equiv r_{\mathrm{horizon}} \; ,
\lab{RN-metric-out}
\ee
where $Q^2 = 8\pi q^2 r_{\mathrm{horizon}}^4 \equiv 128\pi q^2 G^4 M_1^4$;
\end{itemize}
For the Maxwell subsystem we have $\cA_1 = \ldots =\cA_{D-1}=0$ everywhere and:
\begin{itemize}
\item
Coulomb field outside horizon:
\be
\cA_0 = \frac{\sqrt{2}\, q\, r_{\mathrm{horizon}}^2}{r} \;\; ,\quad \mathrm{for}\;\;
r \geq r_0 \equiv r_{\mathrm{horizon}}  \; .
\lab{EM-out}
\ee
\item
No electric field inside horizon:
\be
\cA_0 = \sqrt{2}\, q\, r_{\mathrm{horizon}} = \mathrm{const.} \;\; ,\quad \mathrm{for}\;\;
r \leq r_0 \equiv r_{\mathrm{horizon}}  \; .
\lab{EM-in}
\ee
\end{itemize}

Using the same (synchronous) gauge choice \rf{gauge-fix} and ansatz for the dual
``gauge potential'' \rf{u-ansatz}, as well as taking into account
\rf{EM-out}--\rf{EM-in}, the \textsl{WILL}-membrane equations of motion
\rf{gamma-eqs+A}--\rf{X-eqs+A} acquire the form
(recall $\(\pa_a X \pa_b X\) \equiv \pa_a X^\m \pa_b X^\n G_{\m\n}$):
\be
\(\pa_0 X \pa_0 X\) = 0 \quad ,\quad \(\pa_0 X \pa_i X\) = 0  \; ,
\lab{constr-0+A}
\ee
\be
\(\pa_i X\pa_j X\) - \h \g_{ij} \g^{kl}\(\pa_k X\pa_l X\) = 0 \; ,
\lab{constr-vir+A}
\ee
(these constraints are the same as in the absence of coupling to space-time
gauge field \rf{constr-0}--\rf{constr-vir});
\be
\pa_0 \(\sqrt{\g_{(2)}} \g^{kl}\(\pa_k X\pa_l X\)\) = 0 \; ,
\lab{u-eqs-fix+A}
\ee
(once again the same equation as in the absence of coupling to space-time
gauge field \rf{u-eqs-fix});
\be
{\wti \Box}^{(3)} X^\m + \( - \pa_0 X^\n \pa_0 X^\l +
\g^{kl} \pa_k X^\n \pa_l X^\l \) \G^{\m}_{\n\l}
- q \frac{\g^{kl}\(\pa_k X \pa_l X\)}{\sqrt{2}\,\chi}\,
\pa_0 X^\n \(\pa_\l \cA_\n - \pa_\n \cA_\l \) G^{\l\m} = 0 \; .
\lab{X-eqs-fix+A}
\ee
Here $\chi \equiv T_0 - \sqrt{2}q\cA_0$ with $\cA_0$ as in 
Eqs.\rf{EM-out},\rf{EM-in} is the variable brane tension coming from 
Eqs.\rf{u-ansatz},\rf{tension+A};
$X^0 \equiv t, X^1 \equiv r, X^2 \equiv \th, X^3 \equiv \phi$; and:
\be
{\wti \Box}^{(3)} \equiv
- \frac{1}{\chi \sqrt{\g^{(2)}}} \pa_0 \(\chi \sqrt{\g^{(2)}} \pa_0 \) +
\frac{1}{\chi \sqrt{\g^{(2)}}}\pa_i \(\chi \sqrt{\g^{(2)}} \g^{ij} \pa_j \)
\; .
\lab{box-3+A}
\ee

A self-consistent solution to Eqs.\rf{constr-0+A}--\rf{X-eqs-fix+A} reads: 
\be X^0 \equiv t = \t \quad,\quad \th = \s^1 \quad,\quad \p = \s^2 \;\; ,
\lab{Schwarzschild-RN-sol-1}
\ee 
\be r (\t,\s^1,\s^2) =
r_{\mathrm{horizon}} = \mathrm{const.} \quad,\quad
A_{\pm}(r_{\mathrm{horizon}})=0 \; , 
\lab{Schwarzschild-RN-sol-2}
\ee 
\textsl{i.e.}, the \textsl{WILL}-membrane automatically positions itself on the 
common event horizon of the pertinent black holes. Furthermore, inserting
\rf{Schwarzschild-RN-sol-1}--\rf{Schwarzschild-RN-sol-2} in the expression 
\rf{T-brane} for the \textsl{WILL}-membrane energy-momentum tensor 
$T^{(brane)}_{\m\n}$ and using the simple expressions for the components of
the Ricci tensor corresponding to the metric \rf{spherical-symm-metric-b} 
$R^0_0 = R^1_1 = -\frac{1}{2r^2} \partder{}{r}\( r^2 \partder{}{r}A(r)\)$
\ct{Eduardo-Rabinowitz}, the Einstein equations \rf{Einstein-eqs} entail the 
following important matching conditions for the space-time metric components 
\rf{spherical-symm-metric-b} along the \textsl{WILL}-membrane surface:
\be
\partder{}{r} A_{+}\bgv_{r=r_{\mathrm{horizon}}} -
\partder{}{r} A_{-}\bgv_{r=r_{\mathrm{horizon}}} =
- 16\pi G \chi 
\lab{metric-match} 
\ee
The matching condition \rf{metric-match} corresponds to the so called statically 
soldering conditions in the theory of lightlike thin shell dynamics in general 
relativity in the case of ``horizon straddling'' lightlike matter
(first ref.\ct{Barrabes-Israel-Hooft}). Here, condition \rf{metric-match}
yields relations between the parameters of the black holes and the
\textsl{WILL}-membrane ($q$ being its surface charge density) :
\be 
M_2 = M_1 + 32\pi q^2 G^3 M_1^3 
\lab{mass-match} 
\ee 
and for the brane tension $\chi$: 
\be 
\chi \equiv T_0 - 2q^2r_{\mathrm{horizon}} =q^2 G M_1 \quad, \;\; \mathrm{i.e.} \;\;
T_0 = 5 q^2 G M_1 \lab{tension-match} \ee

We would like to stress that the present {\em WILL}-brane models provide a 
systematic dynamical description of lightlike branes (as sources for both 
gravity and electromagnetism) from first principles starting with concise 
Weyl-conformally invariant actions \rf{WILL-membrane+A}, \rf{E-M-WILL}.
It is interesting that out of the several possibilities discussed in the first 
ref.\ct{Barrabes-Israel-Hooft} for lightlike matter moving in a black hole
gravitational field only the ``horizon straddling'' is selected by the
\textsl{WILL}-branes.

\section{\label{sec:KK}{\em WILL}-Membrane Dynamics in Kaluza-Klein Product
Spaces}

Here we consider {\em WILL}-membrane moving in a general product-space 
$D=(d+2)$-dimensional gravitational background $\cM^d \times \S^{2}$ with 
coordinates $(x^\m, y^m)$ ($\m = 0,1,\ldots ,d-1$, $m = 1,2$). First we take the 
following simple form for the Riemannian metric (which is applicable also to
brane-world scenarios):
\be 
(ds)^2 = f(y) g_{\m\n}(x) dx^\m dx^\n + h_{mn}(y) dy^m dy^n \; . 
\lab{product-metric-0} 
\ee 
The metric $g_{\m\n}(x)$ on $\cM^d$ is of Lorentzian signature. 
Furthermore, we assume that the {\em WILL}-brane wraps around the ``internal''
space $\S^{2}$ and choose the following ansatz (recall $\t \equiv \s^0$) 
which uses the identity mapping between the brane coordinates $\s^1,\s^2$ and 
the coordinates $Y^m$ of the brane:
\be 
X^\m = X^\m (\t) \quad , \quad Y^m = \s^m \quad , \quad
\g_{mn} = a(\t)\, h_{mn}(\s^1,\s^2) \lab{ansatz-product-space-0}
\ee
Then the equations of motion and constraints \rf{constr-0}--\rf{box-3} reduce to: 
\be 
\pa_\t X^\m \pa_\t X^\n g_{\m\n}(X) = 0 \quad ,\quad
\frac{1}{a(\t)}\,\pa_\t \Bigl( a(\t) \pa_\t X^\m \Bigr) + 
\pa_\t X^\n \pa_\t X^\l\, \G^\m_{\n\l} (g) = 0 
\lab{eff-massless} 
\ee
Here $a(\t)$ is the conformal factor of the space-like part of the
internal membrane metric (last Eq.\rf{ansatz-product-space-0}) and
$\G^\m_{\n\l} (g)$ is the Christoffel connection for the
non-compact space metric $g_{\m\n}(x)$. Also, let us note the that
the overall conformal factor $f(u)$ of the metric on $\cM^d$
\rf{product-metric-0} drops out completely.

Eqs.\rf{eff-massless} are of the same form as the equations of
motion for a massless point-particle with a world-line ``einbein''
$e = a^{-1}$ moving in $\cM^d$. In other words, in this
situation we deal with a membrane living in the extra
``internal'' dimensions $\S^{2}$ and moving as a whole with the
speed of light in ``ordinary'' space-time $\cM^d$ -- its motion is
indistinguishable from the dynamics of a regular massless point-particle with respect to the
non-compact projected world $\cM^d$. Notice, however, that although being
massless, the particle-like brane mode acquires non-trivial Kaluza-Klein
quantum numbers due to the \textsl{WILL}-brane winding of the extra compact 
dimensions -- a very peculiar situation in the context of Kaluza-Klein theories.

Now, we take more complicated form for the product-space Riemannian metric: 
\be 
(ds)^2 = g_{\m\n}(x) dx^\m dx^\n + v(x)\,
h_{mn}(y) dy^m dy^n \; , 
\lab{product-metric-1} 
\ee 
where we allow for a variable size (squared) $v(x)$ of the ``internal'' compact
dimensions. Employing the ansatz: 
\be 
X^\m = X^\m (\t,\s^1,\s^2) \quad , \quad Y^m = \s^m \quad , \quad 
\g_{mn} = a(\t)\, h_{mn}(\s^1,\s^2) 
\lab{ansatz-product-space-1} 
\ee 
the constraints and equations of motion \rf{constr-0}--\rf{box-3} acquire the
following form: 
\be 
\(\pa_0 X \pa_0 X\) = 0 \quad ,\quad \(\pa_0 X \pa_i X\) = 0  \quad ,\quad 
\(\pa_i X\pa_j X\) - \h h_{ij} h^{kl}\(\pa_k X\pa_l X\) = 0  \; , 
\lab{constr-KK} 
\ee 
\be
{\widehat \Box}^{(3)} X^\m + \( - \pa_0 X^\n \pa_0 X^\l +
\frac{1}{a} h^{kl} \pa_k X^\n \pa_l X^\l \) \G^{\m}_{\n\l}(g) -
\frac{1}{a} g^{\m\n} \partder{v}{x^\n}\bgv_{x=X} = 0  \; ,
\lab{X-eqs-3-KK} 
\ee 
with: 
\be {\widehat \Box}^{(3)} \equiv - \frac{1}{a} \pa_0 \( a\pa_0 \) + 
\frac{1}{\sqrt{h}}\pa_i \(\sqrt{h} h^{ij} \pa_j \)   \; ; 
\lab{wave-box-3} 
\ee 
\be 
\pa_i X^\m \partder{v}{x^\m}\bgv_{x=X} = 0  \; . 
\lab{Y-eqs-KK} \ee 
Let us also note that Eq.\rf{u-eqs-fix} (the remnant of the equations
of motion for the dual gauge potential $u$ being a consequence of
the constraints and the rest of the equations of motion) upon
using the ansatz \rf{ansatz-product-space-1} assumes the form: 
\be
\pa_0 \( h^{ij}\(\pa_i X \pa_j X\) + 2 v(X)\) = 0  \; .
\lab{u-eqs-KK} 
\ee

In what follows we will study the particle-like mode dynamics of the
{\em WILL}-membrane, \textsl{i.e.}, we will use the ansatz
\rf{ansatz-product-space-0}. Then Eqs.\rf{constr-KK}--\rf{Y-eqs-KK} reduce to:
\be
\pa_0 X^\m g_{\m\n}(X) \pa_0 X^\n = 0 \; ,
\lab{constr-KK-0}
\ee
\be
\frac{1}{a}\pa_0 \( a \pa_0 X^\m\) + \pa_0 X^\n \pa_0 X^\l \G^\m_{\n\l}(g)
+ \frac{1}{a} g^{\m\n} \partder{v}{x^\n}\bgv_{x=X} = 0  \; .
\lab{X-eqs-3-KK-0}
\ee
Let us particularly stress that Eqs.\rf{constr-KK-0}-\rf{X-eqs-3-KK-0} describe 
{\em massless} particle-like dynamics in a ``potential'' $v(X)$ (the
space-dependent size-squared of the extra compact dimensions) which is an 
essential new feature stemming from the {\em WILL}-membrane model. These equations 
{\em cannot} be derived from a reparametrization-invariant (massless) 
point-particle-like action.

For a static spherically-symmetric (with respect to non-compact dimensions) background:
\be
(ds)^2 =  - A(r)(dt)^2 + B(r)(dr)^2 + C(r) \lb (d\th)^2 + \sin^2 (\th)\,(d\p)^2\rb
+ v(r)\, h_{mn}(y) dy^m dy^n \; ,
\lab{spherical-symm-metric-KK}
\ee
and identifying as usual $X^0 \equiv t = \t$ we obtain from
\rf{constr-KK-0}--\rf{X-eqs-3-KK-0} :

\begin{itemize}
\item
As a consequence of Eqs.\rf{constr-KK-0}-\rf{X-eqs-3-KK-0} we have:
\be
\pa_0 X^\m \partder{v}{x^\m}\bgv_{x=X} = 0  \; ,
\lab{u-eqs-KK-0}
\ee
which yields $r=r_0 \equiv \mathrm{const.}$ ($\t$-independent).
\item
The equation of motion \rf{X-eqs-3-KK-0} for $\m = 0$ yields
$a =a_0 \equiv \mathrm{const.}$ ($\t$-independent). Recall from last
Eq.\rf{ansatz-product-space-0} that $a$ has the meaning of a size squared of
the world-surface (at fixed proper-time) of the {\em WILL}-brane with respect to internal
world-volume metric $\g_{ab}$.
\item
Taking the above relations into account, the equation of motion \rf{X-eqs-3-KK-0}
for $\m = r$ yields a purely functional equation:
\be
\(\frac{1}{a_0} \partder{v}{r} - \h A \partder{}{r}\ln\frac{C}{A}\)\bgv_{r=r_0} = 0
\; ,
\lab{r-eqs-KK}
\ee
which determines a set of allowed constant values for $r=r_0 \equiv r(a_0)$
depending on the explicit form of the background \rf{spherical-symm-metric-KK}
and parametrically depending on $a_0$.
\item
Finally, the lightlike constraint (first Eq.\rf{constr-KK-0}) and the
equations of motion \rf{X-eqs-3-KK-0} for the space-like non-compact coordinates
$\underline{X}\equiv \( X^1, X^2, X^3 \)$ (recall
$r \equiv \sqrt{(X^1)^2 + (X^2)^2 + (X^3)^2} = r_0$) acquire the form:
\be
{\stackrel{.}{\underline{X}}}^2 = \frac{A}{C}\, r^2\bgv_{r=r_0} \quad ,\quad
\stackrel{..}{\underline{X}} + \frac{A}{C}\bgv_{r=r_0} \underline{X} = 0
\lab{X-eqs-KK}
\ee
Here and below the scalar products of 3-vectors are understood with respect to flat
metric.
\end{itemize}

Obviously, the solution to \rf{X-eqs-KK} are as follows:
\be
\underline{X} = r_0 \( \underline{n}^{(1)} \cos (\om \t) +
\underline{n}^{(2)} \sin (\om \t)\)  \quad ,\quad
\om^2 \equiv \frac{A}{C}\bgv_{r=r_0}
\lab{X-sol}
\ee
where $r_0$ is determined from Eq.\rf{r-eqs-KK} and $\underline{n}^{(1,2)}$
are two constant mutually orthogonal unit 3-vectors:
$\(\underline{n}^{(1)}\)^2 = \(\underline{n}^{(2)}\)^2 = 1$,
$\underline{n}^{(1)}.\underline{n}^{(2)} = 0$. In other words, the
projection of the {\em WILL}-membrane on the non-compact space-time rotates
with the speed of light in a two-dimensional plane defined by the unit vectors
$\underline{n}^{(1,2)}$, traversing a circle with radius $r_0$ determined from
Eq.\rf{r-eqs-KK} with angular velocity $\om$ given by the second relation
\rf{X-sol}.

Solution \rf{X-sol} describes nontrivial massless mode dynamics with energy
$E$ and angular momentum $|\underline{M}|$ (recall $\chi = T_0$ \rf{chi-0}) :
\be
E = T_0 a_0 A (r_0) \Omega \quad , \quad 
|\underline{M}| = T_0 a_0 \sqrt{A (r_0) C(r_0)}\, \Omega  \; ,
\lab{energy-angular-mom}
\ee
$\Omega$ being the volume of the compact ``internal'' space, which
implies the relation:
\be
E = |\underline{M}|\,\om (r_0)  \; .
\lab{E-M-rel}
\ee
(recall $\om (r_0) \equiv \sqrt{A (r_0) / C(r_0)})$ \rf{X-sol}). 
Eqs.\rf{energy-angular-mom} follow straightforwardly from the expressions
for the Noether conserved currents corresponding to invariance of the action 
\rf{WILL-membrane} with external metric background given by 
\rf{spherical-symm-metric-KK} under translation of
$X^0 \equiv t$ and $\p$ (external space-time spherical angle).

%

Following refs.\ct{aaron-dowen,yoshimura}, we have the following explicit solution of the
Einstein equations in the case of two extra dimensions for the metric 
\rf{spherical-symm-metric-KK} :
\be
A (r) = \(\frac{\a r - 1}{\a r + 1}\)^{2p + q} \;\; ,\;\;
C (r) = r^2 B(r) = \frac{1}{\a^4 r^2}\, 
\frac{(\a r + 1)^{2(p+1)-q}}{(\a r - 1)^{2(p-1)-q}} \;\; ,\;\;
v (r) = \(\frac{\a r + 1}{\a r - 1}\)^q
\lab{6D-sol}
\ee
whereas the compact internal space $\S^2$ is a torus with $h_{mn}(y) = \d_{mn}$
(therefore, the world-surface of the {\em WILL}-membrane is similarly assumed to 
have toroidal topology). In \rf{6D-sol} $\a$ is arbitrary positive integration 
constant of mass dimension 1, $p$ and $q$ are free numerical parameters subject to the 
relation $p^2 +\h q^2 =1$, and the resulting metric is well-defined in the region 
$\a r > 1$. 

Inserting the expressions \rf{6D-sol} into Eq.\rf{r-eqs-KK} we obtain:
\be
0 = \frac{4 q}{a_0}\,\frac{(\a r_0 + 1)^{q-1}}{(\a r_0 - 1)^{q+1}} +
\(\frac{\a r_0 - 1}{\a r_0 + 1}\)^{p - q}
\biggl\lbrack \frac{(4p+2)}{\a r_0 + 1} - \frac{2}{\a r_0} - 
\frac{(4p-2)}{\a r_0 - 1}\biggr\rbrack
\lab{r-eqs-KK-6D-sol}
\ee
determining allowed values of the radius $r_0 = r_0 (a_0)$ of the planar circular orbits
as a function of $a_0$ (recall that $a_0$ is strictly positive free parameter -- 
the size squared of the \textsl{WILL}-membrane with respect to its internal world-volume metric,
cf. Eq.\rf{ansatz-product-space-0}). From Eq.\rf{r-eqs-KK-6D-sol} and taking into 
account the constraints on the parameters in \rf{6D-sol}, we have:
\begin{itemize}
\item
For $0 < q \leq 1/\sqrt{2}$ and $\h \leq p \leq 1 - \h q^2$ the allowed values of
the radius $r_0 (a_0)$ of the planar circular orbits lie in the interval 
$(\frac{1}{\a}\, ,\, \frac{2}{\a}\( p + \sqrt{p^2 - 1/4}\)$.
\item
For $-1/\sqrt{2} \leq q < 0$ and $\h \leq p \leq 1 - \h q^2$ the allowed values of 
$r_0 (a_0)$ are $r_0 (a_0) \geq \frac{2}{\a}\( p + \sqrt{p^2 - 1/4}\)$; we have in 
this case $r_0 (a_0) \longrightarrow \infty$ for $a_0 \to 0$, and 
$r_0 (a_0) \longrightarrow \frac{2}{\a}\( p + \sqrt{p^2 - 1/4}\)$ for $a_0 \to \infty$.

\end{itemize}

Let us note that in the case $q>0$ \rf{6D-sol} also yields the interior solution
for the ``gravitational bags'' \ct{grav-bag} (the latter similarly require
$p>\h$) \footnote{For ``gravitational bags'' \ct{grav-bag} the exterior solution, 
matched to the interior one through a regular domain wall, is a Schwarzschild metric.}.

For both situations above one can easily show that the allowed values for
the angular velocity (cf. the definition in Eq.\rf{X-sol}) :
\be
\om (r_0) = \a^2 r_0 \frac{\(\a r_0 - 1\)^{2p-1}}{\(\a r_0 + 1\)^{2p+1}}
\lab{omega-eq}
\ee
lie in a finite interval between 0 and $\om_{\mathrm{max}}$:
\be
\om_{\mathrm{max}} = \frac{\a}{2}\,\frac{\(p+\sqrt{p^2 - 1/4}\)
\(p-1/2 +\sqrt{p^2 - 1/4}\)^{2p-1}}{\(p+1/2 +\sqrt{p^2 - 1/4}\)^{2p+1}}
\lab{omega-max}
\ee

Let us note that in the case $v = \mathrm{const.}$ (constant size of extra
dimensions) the free scale parameter $a_0$ disappears in Eq.\rf{r-eqs-KK} leading to
a qualitatively different situation. For instance, in the case of Schwarzschild
metric on the non-compact space, \textsl{i.e.}, $A(r)=B^{-1}(r) = 1 - 2GM/r
\; ,\; C(r)=r^2 \;, \; v = \mathrm{const.}$ in \rf{spherical-symm-metric-KK},
Eq.\rf{r-eqs-KK} is satisfied for only two special values of $r$ \ct{MTW}:
$r_0 = 2GM$ (massless particle ``sitting'' on the horizon) and $r_0 = 3GM$
(massless particle on an unstable circular orbit). On the other hand, for variable 
size of the extra dimensions ($\partder{v}{r}\neq 0$) a continuous range of
values for the ``radius'' of the planar circular orbits is available corresponding 
to the solutions $r=r_0 (a_0)$ of Eq.\rf{r-eqs-KK}.

Let us particularly emphasize the fact that, although the {\em WILL}-brane
is wrapping the extra (compact) dimensions in a topologically non-trivial way
(cf. second Eq.\rf{ansatz-product-space-1}), its modes remain {\em massless} from the
projected non-compact space-time point of view. This is a new phenomenon
from the point of view of Kaluza-Klein theories: here we have particle-like
membrane modes, which acquire non-zero quantum numbers due to non-trivial winding,
while at the same time these particle-like modes remain massless. In contrast, 
one should recall that in ordinary Kaluza-Klein theory (for reviews, 
see \ct{K-K-review}), non-trivial dependence on the extra dimensions is possible 
for point particles or even standard strings and branes only at a very high
energy cost (either by momentum modes or winding modes), which implies a
very high mass from the projected non-compact space-time point of view.

\section{\label{sec:outlook}Conclusions and Outlook}

In the present paper we have discussed in detail a completely new type of
$p$-branes. The use of a modified non-Riemannian volume form (integration
measure) in their Lagrangian actions was of crucial importance. Next,
formulating acceptable $p$-brane dynamics naturally requires the introduction 
of additional world-volume gauge fields. Employing a square-root Maxwell-type 
action for auxiliary world-volume gauge field was most instrumental for achieving
a consistent $p$-brane theory which is manifestly Weyl-conformally invariant for 
any $p$ and, furthermore, for any even $p$ (odd-dimensional world-volume) it 
describes intrinsically lightlike $p$-branes. Remarkably, the brane tension 
becomes now a gauge-dependent concept -- it appears as a composite field
transforming non-trivially under Weyl-conformal transformations. Unlike
previous Weyl-invariant {\em reformulations} of the standard Weyl
non-invariant Nambu-Goto $p$-branes (which preserve the physical content of
the Nambu-Goto branes and, therefore, describe massive brane modes), the
presently discussed new class of Weyl-invariant $p$-branes for
$p+1=\mathrm{odd}$ describes genuine massless lightlike branes.

Weyl-invariant lightlike $p$-branes (\textsl{WILL}-branes) offer a broad
variety of interesting physical applications, most notably in the context of
black hole physics and Kaluza-Klein theories. In the case of a \textsl{WILL}-membrane
moving as a test-brane in a gravitational black hole background we have seen
that it positions itself automatically on the event horizon. Furthermore, we
studied a self-consistent solution of the coupled Einstein-Maxwell-\textsl{WILL}-membrane
system where the \textsl{WILL}-membrane appears as a source for both gravity
and electromagnetism. This self-consistent solution has the \textsl{WILL}-membrane
``sitting'' at (``straddling'') the common event horizon of a Schwarzschild 
(in the interior) and Reissner-Nordstr\"{o}m (in the exterior) black hole solutions. 
This is an indication that the \textsl{WILL}-membrane model indeed provides a plausible
explicit dynamical realization of the so called ``membrane paradigm'' in
black hole physics \ct{membrane-paradigm}.
The quantization of the \textsl{WILL}-membrane dynamics under these
circumstances may be very much related to the quantization of the horizon
degrees of freedom. Indeed, the \textsl{WILL}-membrane presents a remarkable
resemblance to the string-like objects introduced by `t Hooft \ct{Hooft}
to characterize the horizon degrees of freedom.

In the context of Kaluza-Klein theories the \textsl{WILL}-branes appear also
to play a very interesting role. Indeed, we have found solutions for the 
\textsl{WILL}-membrane moving in higher-dimensional Kaluza-Klein-type
space-times which describe the dynamics of {\em massless} particle-like
brane modes even though the membrane itself is wrapping the extra compact
dimensions and, therefore, acquires non-trivial Kaluza-Klein charges -- a
situation inaccessible in the context of standard Kaluza-Klein theories.
When the size of the extra compact dimensions has a nontrivial space
dependence like in some self-consistent solutions of higher-dimensional
Einstein equations \ct{aaron-dowen,yoshimura}, the behavior of the massless
particle-like brane mode solutions is quite interesting from the point of
view of the non-compact $D=4$ space-time point of view. These massless brane
modes are trapped on finite planar circular orbits with linear dependence 
between energy and angular momentum. The parameters of the metric on the
non-compact part of the Kaluza-Klein space-time dictate that the allowed
values of the angular velocity lie in a finite interval. 

It is essential to note that the above massless particle-like dynamics is a special 
feature due to its \textsl{WILL}-brane \rf{WILL-membrane-0} origin. It cannot be derived 
neither from a reparametrization-invariant point-particle action nor by zero-mode
reduction of a Nambu-Goto-type brane action.

There are various physically interesting directions for further systematic study 
of the properties and implications of the new class of Weyl-conformally
invariant branes discussed above, such as:
quantization (Weyl-conformal anomaly and critical dimensions); supersymmetric
generalization; possible relevance for the open string dynamics (similar to the
role played by Dirichlet- ($Dp$-)branes); {\em WILL}-brane dynamics in more 
complicated gravitational black hole backgrounds (\textsl{e.g.}, Kerr-Newman);
{\em WILL}-brane dynamics in more complicated Kaluza-Klein-type space-times,
including more complex winding of the extra dimensions. To this end let us
note that there exist physically interesting solutions of higher-dimensional
Einstein equations -- ``gravitational bags'' \ct{grav-bag} and ``dimension bubble''
solutions \ct{bubble}, where the presence of a domain wall implies big gradient 
for the size-squared $v(r)$ of the extra dimensions (cf. Eq.\rf{r-eqs-KK}. 
Thus, it would be very interesting to study \textsl{WILL}-brane dynamics in 
such Kaluza-Klein backgrounds.

\textbf{Acknowledgements.}
{\small Two of us (E.N. and S.P.) are sincerely grateful for hospitality and
support to the organizers of
the \textsl{Third Summer School on Modern Mathematical Physics}
(Zlatibor, Serbia and Montenegro, August 2004), and the
\textsl{Second Annual Meeting of the European RTN}~ {\em EUCLID}
(Sozopol, Bulgaria, September 2004), where the above results were
first presented. One of us (E.G.) thanks the Institute for Nuclear
Research and Nuclear Energy (Sofia) and Trieste University for
hospitality. He also acknowledges useful conversations with
Stefano Ansoldi, Carlos Castro, Gerard `t Hooft, Alexander Polyakov and Euro
Spallucci.

E.N. and S.P. are partially supported by Bulgarian NSF grant \textsl{F-1412/04}
and European RTN {\em ``Forces-Universe''} 
(Contract No.\textsl{MRTN-CT-2004-005104}).
Finally, all of us acknowledge support of our collaboration through the exchange
agreement between the Ben-Gurion University of the Negev (Beer-Sheva, Israel) and
the Bulgarian Academy of Sciences.}

\end{document}